\documentclass[%
 reprint,
superscriptaddress,
 amsmath,amssymb,
 aps,
]{revtex4-2}

\usepackage{graphicx}
\usepackage{dcolumn}
\usepackage{bm}
\usepackage{comment}
\usepackage{amsmath}
\usepackage{physics}
\usepackage{bm}
\usepackage{subfigmat}
\usepackage{mathrsfs}
\usepackage{array}

\begin{document}

\preprint{APS/123-QED}

\title{Scalable Conflict-free  Decision Making with Photons }

\author{Kohei Konaka}
 
 \affiliation{
 Graduate School of Information Science and Technology, The University of Tokyo, 7-3-1 Hongo, Bunkyo-ku, Tokyo 113-8656, Japan}
\author{André Röhm}
 \affiliation{
 Graduate School of Information Science and Technology, The University of Tokyo, 7-3-1 Hongo, Bunkyo-ku, Tokyo 113-8656, Japan}
\author{Takatomo Mihana}
\email{takatomo\_mihana@ipc.i.u-tokyo.ac.jp}
 \affiliation{
 Graduate School of Information Science and Technology, The University of Tokyo, 7-3-1 Hongo, Bunkyo-ku, Tokyo 113-8656, Japan}
\author{Ryoichi Horisaki}
 \affiliation{
 Graduate School of Information Science and Technology, The University of Tokyo, 7-3-1 Hongo, Bunkyo-ku, Tokyo 113-8656, Japan}

\date{\today}

\begin{abstract}
Quantum optics utilizes the unique properties of light for computation or communication. In this work, we explore its ability to solve certain reinforcement learning tasks, with a particular view towards the scalability of the approach. Our method utilizes the Orbital Angular Momentum (OAM) of photons to solve the Competitive Multi-Armed Bandit (CMAB) problem while maximizing rewards.   In particular, we encode each player's preferences in the OAM amplitudes, while the phases are optimized to avoid conflicts. We find that the proposed system is capable of solving the CMAB problem with a scalable number of options and demonstrates improved performance over existing techniques. As an example of a system with simple rules for solving complex tasks, our OAM-based method adds to the repertoire of functionality of quantum optics.
\end{abstract}

\maketitle


\section{Introduction}
\begin{figure*}[tb]
    \centering
    \includegraphics[width=0.90\linewidth,keepaspectratio]{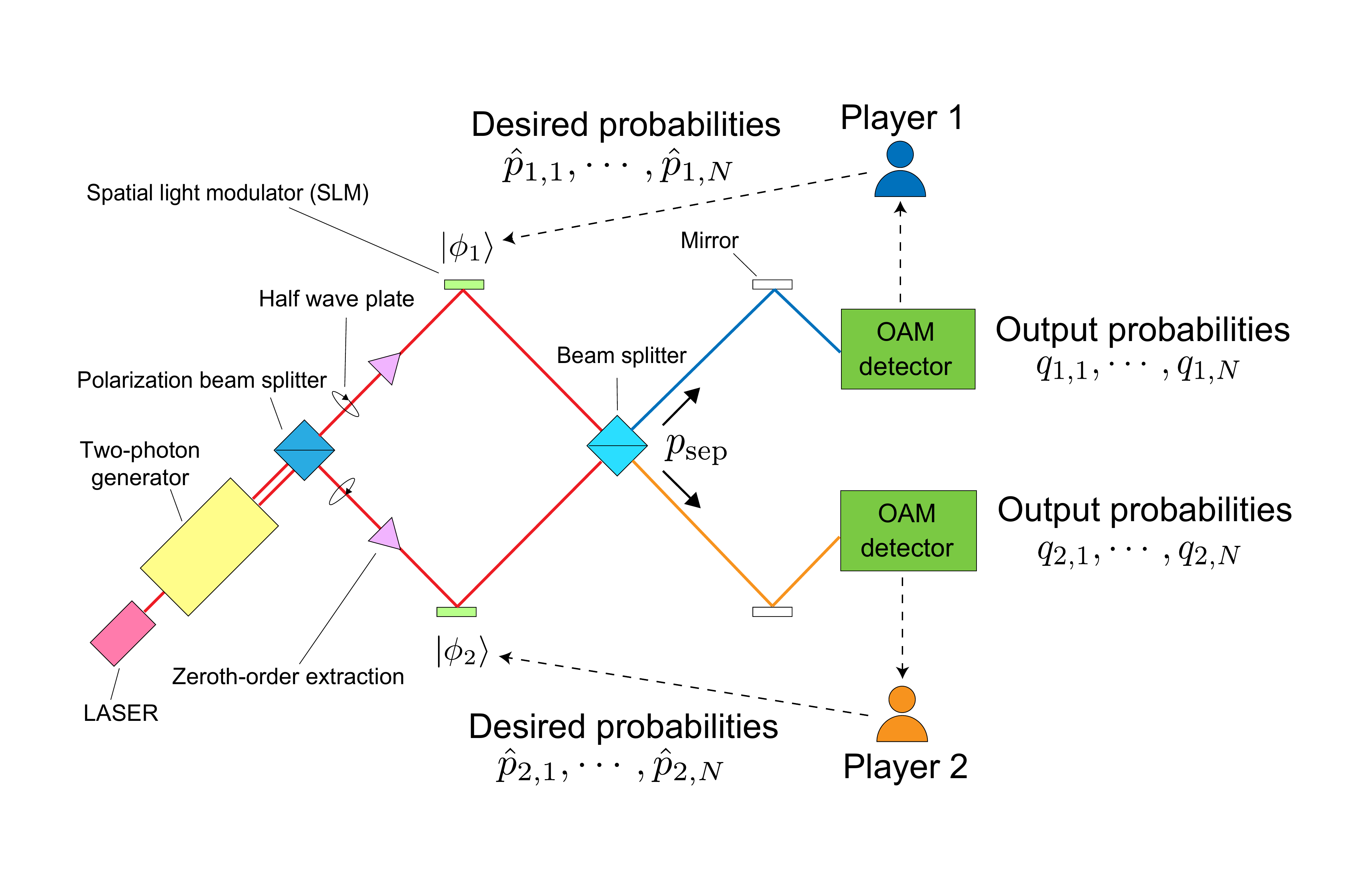}
    \caption{Schematic diagram of the proposed decision making system. First, a polarization-entangled photon pair is created by the two-photon generator and then each photon is sent to SLM 1 and SLM 2, respectively, by the polarization beam splitter. The two photons are changed to specific OAM states by the SLMs according to Eqs. \eqref{OAM_1} and \eqref{OAM_2}. The probability amplitudes of the OAM states encode the desired probabilities of the players, and the phases of the OAM states are determined by the method explained in Sec. \ref{subsec:optimization}. The two photons undergo quantum interference on a beam splitter and their OAM states are subsequently measured by each player. The SLMs are then updated based on the received reward.}
    \label{proposed-fig}
\end{figure*}
We are constantly forced to make decisions with limited information in the real world. 
Yet, we can also learn from our experience to avoid bad choices and prefer those that have previously led to good outcomes. 
Reinforcement learning \cite{sutton1999reinforcement} is the framework that models such learning steps, in particular in stochastic environments with unknown reward distributions.
Among the models of reinforcement learning, the Multi-Armed Bandit (MAB) problem \cite{MAB} is arguably the clearest framework for illustrating the central tension of this class of tasks.

In the MAB problem, we consider a finite number of options, called arms, which generate rewards according to certain probability distributions when selected.
The basic scenario involves a single player who repeatedly chooses one arm from multiple options over a fixed number of trials, aiming to maximize the cumulative rewards.
The player does not know the probability distributions from which the arms generate rewards.
Therefore, the player must first perform ``exploration," selecting each arm at least some times to identify those with higher expected rewards.
However, to maximize cumulative rewards, the player also needs to perform ``exploitation," focusing on selecting the arm with higher expected rewards.
Balancing exploration and exploitation effectively is crucial and the core aspect modeled by the MAB problem \cite{Exploration_and_exploitation}. 
Due to the generality of this model, the MAB problem has been applied in various fields, ranging from online advertising
optimization \cite{advertisement1,advertisement2} to clinical trials of new medicines \cite{drug1,drug2,drug3}.
Well-known algorithms for solving the MAB problem include the Softmax method \cite{softmax}, Thompson sampling \cite{Thompson_sampling}, and the Upper confidence bound method \cite{UCB}.

One extension of the MAB problem involves considering multiple players, particularly addressing the issue of reward division when conflicts in selection (referred to as selection conflicts) occur. This problem is known as the Competitive Multi-Armed Bandit (CMAB) problem \cite{CMAB1,CMAB2,CMAB3,CMAB4}. 
In the CMAB problem, the goal is to maximize the sum of cumulative rewards across all players. Therefore, each player must engage in both ``exploration" and ``exploitation" while avoiding selection conflicts. 
The CMAB problem is relevant to various applications such as frequency allocation in wireless communications \cite{cognitive_radio}, where selection conflicts result in performance degradation of individual devices due to multiple devices using the same frequency band. 
To avoid selection conflicts, direct communication between players (users of wireless communications) is often undesirable from the perspective of time and energy efficiency. 
Consequently, a common assumption in the CMAB problem, which this paper also adopts, is that players cannot communicate to share selection information (which arms players select and whether the selected arms generate rewards) directly. 
In naive extensions of algorithms used for the MAB problem it is extremely difficult to avoid selection conflicts without sharing selection information.

Recent studies have been conducted to solve the CMAB problem through physical decision making using the properties of light. 
For example, decision making that avoids selection conflicts through quantum interference of photons has been achieved by utilizing properties like the polarization state and the Orbital Angular Momentum (OAM) of light \cite{OAM,Amakasu}. OAMs are spatial field distributions characterized by helical phase fronts and quantized angular momentum; they have the potential to encode a theoretically infinite number of states, unlike polarization states.
In practice, OAM modes more than $l=50$ have been demonstrated \cite{lavery2013efficient}.
Additionally, some studies have employed laser networks, consisting of multiple lasers exhibiting chaotic synchronization, to address the problem \cite{laser1,laser2}.
Utilizing the properties of light and the performance of optical systems holds promise for creating new system architectures leveraging the unique features of light, potentially surpassing traditional computational systems.

In this work, we present an approach based on a quantum optical implementation using OAM.  
The method previously introduced by Amakasu \textit{et al.}\cite{Amakasu} was mainly based on attenuators. In short, each player was essentially post-selecting photons, by varying the absorption rates proportional to their desired probability ratios. However, this approach greatly limits the number of applicable arms to a maximum of $N=4$. 
Furthermore, the previous method showed  a strong performance dependence on the assignment of arm indices.
Therefore, we propose directly encoding the estimation of players into the OAM states of photons, instead of any attenuators.
We theoretically investigate the scaling properties of our proposed method with regard to two players and an arbitrary number of arms. 
We find that the proposed system is effective in addressing the CMAB problem with a greater number of options and exhibits enhanced performance.

\section{Proposed system}\label{Proposed_method_section}
\subsection{Formulation of the MAB and CMAB problem}
First, we mathematically define the MAB problem, where there is a single player. 
The player is given $N$ arms, and at each time step $t=1,\,\cdots,\,T$, if the player selects Arm $n\, (1\leq n \leq N)$, it generates the reward $r_{n}(t)$ stochastically.
Hereinafter, for the sake of simplicity, let the reward $r_{n}(t)$ follow the Bernoulli distribution $B(1,\mu_{n})$:
\begin{align}
    r_{n}(t)=
    \begin{cases}
        1 & (\text{with probability $\mu_{n}$})\\
        0 & (\text{with probability $1-\mu_{n}$})
    \end{cases}.
\end{align}
In addition, we assume that the reward probability $\mu_{n}$ is unknown to the player.
In the MAB problem, the player's goal is to maximize the cumulative reward obtained from $T$ choices.  
Then, there is a trade-off between ``exploration" (identifying the best arm) and ``exploitation" (focusing on the arm which is currently estimated to have the highest reward probability).

Next, we consider the CMAB problem, where there are $M(\geq 2)$ players. $M$ players repeatedly choose one out of $N$ arms. A key distinction from the simple MAB problem is that when multiple players choose the same arm, the reward is shared equally among them. Then, the reward is defined as
\begin{align}
    r_{n}(t) = \begin{cases}
        \frac{1}{\gamma_{n}(t)} & (\text{with probability $\mu_{n}$})\\
        0 & (\text{with probability $1-\mu_{n}$})
    \end{cases},
\end{align} 
where $\gamma_{n}(t)$ is the number of players choosing  Arm $n$ at time step $t$. Therefore, in addition to the trade-off between ``exploration" and ``exploitation," it becomes crucial to avoid selection conflicts.
\subsection{Encoding desired probabilities into OAM states}

We propose a system consisting of a two-photon generator, a polarization beam splitter, two SLMs, a beam splitter and detectors for solving the CMAB problem (see FIG.~\ref{proposed-fig}).
Here, each SLM is configured based on the current best knowledge of one of the players. 
We use an encoding based on the Orbital Angular Momentum (OAM) modes.
Let $\ket{l}$ describe the quantum state in which there is one photon in the $l$th OAM mode. 
The OAM state of a photon can be expressed by a linear superposition with $\ket{l}$ as the orthonormal basis.

The proposed system is illustrated in FIG. \ref{proposed-fig}.
First, a polarization-entangled photon pair is created by the two-photon generator and then each photon is sent to SLM 1 and SLM 2, respectively, by the polarization beam splitter. 
Player 1 controls SLM 1, and Player 2 controls SLM 2, thereby imparting the following OAM states to the photons:
\begin{align}
    \ket{\phi_{1}}&=\sum_{n=1}^{N}\sqrt{\hat{p}_{1,n}}\mathrm{e}^{\mathrm{i}\theta_{1,n}}\ket{n},\label{OAM_1}\\
    \ket{\phi_{2}}&=\sum_{n=1}^{N}\sqrt{\hat{p}_{2,n}}\mathrm{e}^{\mathrm{i}\theta_{2,n}}\ket{-n},\label{OAM_2}
\end{align}
where $\theta_{m,n}\in\mathbb{R}$ is the phase, and $\hat{p}_{m,n}$ is the desired probability, calculated from the history
of the chosen arms and the rewards obtained. 
In this method, the desired probabilities are encoded into the probability amplitudes of the OAM states of photons. 
The desired probabilities $\hat{p}_{m,n}$ of player $m\in\{1,2\}$ for arm $n\in\{1,\dots,N\}$ can be expressed in a generalized form \cite{Amakasu}:
\begin{align}
    \hat{p}_{m,n} = \frac{1}{2}\frac{\mathrm{e}^{\beta_{m} \hat{\mu}_{m,n}}}{\sum_{k=1}^{N}\mathrm{e}^{\beta_{m} \hat{\mu}_{m,k}}}\left(1+\sum_{n^{\prime}\neq n}\frac{\mathrm{e}^{\beta_{m} \hat{\mu}_{m,n^{\prime}}}}{\sum_{k\neq n^{\prime}}\mathrm{e}^{\beta_{m} \hat{\mu}_{m,k}}}\right), \label{proposed_calculate_sel_prob}
\end{align}
where $\hat{\mu}_{m,n}$ is the empirical reward probability for arm $n$ observed by player $m$.
This is an extension of a classical algorithm known as the Softmax method \cite{softmax} for the MAB problem, where $\beta_{m}$ is a hyperparameter referred to as the inverse temperature parameter.
Generally, the smaller $\beta$, the more the decision making emphasizes exploration. 
In contrast, the larger $\beta$, the more it emphasizes exploitation. 
In the CMAB, it is generally useful to transition from exploration to exploitation as time progresses.
Therefore, the performance of the Softmax method can be improved by increasing $\beta_{1}(t)$ and $\beta_{2}(t)$ over time.
We choose a simple linear scaling with $\beta_{1}(t)=\lambda_{1}t$ and $\beta_{2}(t)=\lambda_{2}t$ \cite{softmax_parameter}. 
Although $\lambda_{1}$ and $\lambda_{2}$ can generally differ, we assume $\lambda=\lambda_{1}=\lambda_{2}$ for simplicity in this paper.

How much each player wants to play each arm depends on their history of wins and losses, which determines $\hat{\mu}_{m,n}$. 
This is then encoded in the amplitudes.
Conversely, the phases $\theta_{m,n}$ are decided by the method explained in Sec. \ref{subsec:optimization}. 

The two photons with imparted OAM states undergo quantum interference at the beam splitter, and if the two photons exit at different output ports, the OAM of each photon is detected.
Methods for observing OAMs include the use of computer-generated holograms with single-mode fibers and the use of interferometers, among others. 
When the player observes the value of the OAM, the player selects the arm corresponding to the absolute value of the OAM.
That selection may yield reward, or it may not, which will change $\hat{\mu}_{m,n}$.
The desired probabilities $\hat{p}_{m,n}$ are then updated, and the process is repeated.

\subsection{Formulation of output probabilities}
In the proposed method, each player calculates the desired probabilities $\hat{p}_{m,n}$ from the history of the chosen arms and the rewards obtained and encodes them into the OAM states. 
However, the probability that each player observes their respective OAM values, i.e., the probability of selecting each arm, is generally expected to differ from the desired probability encoded by each player. 
To distinguish this probability from the desired probability encoded into the OAM states, we refer to it as the output probability. 
The difference between desired and output probabilities mainly stems from the fact that conflict-avoidance puts additional constraints on the arm selection.
In this section, we derive  the output probabilities.

Here, we consider the case where a photon in the OAM state $\ket{\phi_{1}}$ is incident at the first input, and a photon in the OAM state $\ket{\phi_{2}}$ is incident at the second input port, with OAMs being measured at the first and second output subsequently. 
Let $\ket{\psi,k}$ represent the state where a photon in the OAM state $\ket{\psi}$ is at the $k$-th input/output port ($k=1,\,2$). In the optimal circuit of the proposed method depicted in FIG. \ref{proposed-fig}, the input OAM states of photons are transformed as follows:
\begin{align}
    \ket{\phi_{1},1}&\rightarrow -\frac{1}{\sqrt{2}}\ket{\phi_{1},1}+\frac{\mathrm{i}}{\sqrt{2}}\ket{\mathscr{R}\phi_{1},2}\label{OAM_trans_1}\\
    \ket{\phi_{2},2}&\rightarrow \frac{\mathrm{i}}{\sqrt{2}}\ket{\mathscr{R}\phi_{2},1}-\frac{1}{\sqrt{2}}\ket{\phi_{2},2},\label{OAM_trans_2}
\end{align}
where $\mathscr{R}$ represents an operator that inverts the OAM when a photon is reflected by a mirror or beam splitter;
\begin{align}
    \mathscr{R}\left(\sum_{-\infty}^{\infty}\alpha_{l}\ket{l}\right)=\sum_{-\infty}^{\infty}\alpha_{l}\ket{-l}.
    \label{reflection}
\end{align}
The derivation of Eqs. \eqref{OAM_trans_1} and \eqref{OAM_trans_2} is explained in Appendix \ref{Appendix_beamsplitter}.
Thus, the state of the two photons at the output is expressed by
\begin{align}
    \frac{-\ket{\phi_{1},1}+\mathrm{i}\ket{\mathscr{R}\phi_{1},2}}{\sqrt{2}}\otimes_{\mathrm{s}}\frac{\mathrm{i}\ket{\mathscr{R}\phi_{2},1}-\ket{\phi_{2},2}}{\sqrt{2}}.\label{two-photon state}
\end{align}
Here $\otimes_{\mathrm{s}}$ denotes the symmetric tensor product, which for quantum states $\ket{a},\ket{b}$ is defined as:
\begin{align}
    \ket{a}\otimes_{\mathrm{s}}\ket{b} = \frac{1}{\sqrt{2}}\left(\ket{a}\otimes\ket{b}+\ket{b}\otimes\ket{a}\right)
\end{align}
where $\otimes$ is the tensor product.

When we expand Eq. \eqref{two-photon state}, let  $\ket{\mathcal{E}_{1;1}}$ be the state, where the two photons are in separate output ports: 
\begin{align}
    \ket{\mathcal{E}_{1;1}}&=\frac{1}{2}\ket{\phi_{1},1}\otimes_{\mathrm{s}}\ket{\phi_{2},2}-\frac{1}{2}\ket{\mathscr{R}\phi_{1},2}\otimes_{\mathrm{s}}\ket{\mathscr{R}\phi_{2},1}\nonumber\\
    \begin{split}
        &=\frac{1}{2}\left(\sum_{n=1}^{N}c_{1n}\ket{n,1}\right)\otimes_{\mathrm{s}}\left(\sum_{n=1}^{N}c_{2n}\ket{-n,2}\right)\\
        & \quad -\frac{1}{2}\left(\sum_{n=1}^{N}c_{2n}\ket{n,1}\right)\otimes_{\mathrm{s}}\left(\sum_{n=1}^{N}c_{1n}\ket{-n,2}\right),
    \end{split}
    \label{separate_state}
\end{align}
where 
\begin{align}
    c_{mn} = \sqrt{\hat{p}_{m,n}}\mathrm{e}^{\mathrm{i}\theta_{m,n}}.
\end{align}
Photons do not always end up in separate output ports, as can be seen from the fact that Eq.~\eqref{separate_state} is not the complete state.
Let $p_{\mathrm{sep}}$ denote the probability that two photons are observed at separate output ports.
It is given by:
\begin{align}
    p_{\mathrm{sep}}&=\left|\braket{\mathcal{E}_{1;1}}\right|^{2}\nonumber\\
    &=|\frac{1}{2}|^{2}+|\frac{1}{2}|^{2}-2\cdot |\frac{1}{2}|^{2}\left|\braket{\phi_{1}}{\mathscr{R}\phi_{2}}\right|^2\nonumber\\
    &=\frac{1}{2}-\frac{1}{2}\left|\braket{\phi_{1}}{\mathscr{R}\phi_{2}}\right|^2,
    \label{general_p_separate}
\end{align}
where we use the following fact which can be verified through the simple calculation:
\begin{align}
    &\quad (\bra{\phi_{1},1}\otimes_{\mathrm{s}}\bra{\phi_{2},2})(\ket{\mathscr{R}\phi_{1},2}\otimes_{\mathrm{s}}\ket{\mathscr{R}\phi_{2},1})\nonumber\\
    \begin{split}
        &=\frac{1}{\sqrt{2}}(\bra{\phi_{1},1}\otimes\bra{\phi_{2},2}+\bra{\phi_{2},2}\otimes\bra{\phi_{1},1})\\
        &\quad \cdot \frac{1}{\sqrt{2}}(\ket{\mathscr{R}\phi_{1},2}\otimes\ket{\mathscr{R}\phi_{2},1}+\ket{\mathscr{R}\phi_{2},1}\otimes\ket{\mathscr{R}\phi_{1},2})
    \end{split}\nonumber\\
    &=\left|\braket{\phi_{1}}{\mathscr{R}\phi_{2}}\right|^2.
\end{align}
Moreover, the probability $\mathrm{Pr}\left(n_{1},-n_{2}\right)$ of observing the two photons in the first output port with OAM $+n_{1}$ and in the second output port with OAM $-n_{2}$ is given by the square of the absolute value of the coefficient of $\ket{n_{1},1}\otimes_{\mathrm{s}}\ket{-n_{2},2}$ in Eq. \eqref{separate_state}, which means
\begin{align}
    &\quad \mathrm{Pr}\left(n_{1},-n_{2}\right)\nonumber\\
    &=\frac{1}{4}\left|c_{1,n_{1}}c_{2,n_{2}}-c_{1,n_{2}}c_{2,n_{1}}\right|^2\nonumber\\
    \begin{split}
        &=\frac{1}{4}\biggl(\hat{p}_{1,n_{1}}\hat{p}_{2,n_{2}}+\hat{p}_{1,n_{2}}\hat{p}_{2,n_{1}} \\
        & \quad -2\sqrt{\hat{p}_{1,n_{1}}\hat{p}_{1,n_{2}}\hat{p}_{2,n_{1}}\hat{p}_{2,n_{2}}}\cos (\omega_{n_{2}}-\omega_{n_{1}})\biggr),
    \end{split}
\end{align}
where 
\begin{align}
    \omega_{n} = \theta_{2,n}-\theta_{1,n}\ (n=1,\cdots,N).
\end{align}
If $n_{1}=n_{2}=n$, 
\begin{align}
    \mathrm{Pr}\left(n_{1},-n_{2}\right)=\frac{1}{4}\left(\hat{p}_{1,n}\hat{p}_{2,n}+\hat{p}_{1,n}\hat{p}_{2,n}-2\hat{p}_{1,n}\hat{p}_{2,n}\right)=0.
\end{align}
Therefore, the selection conflicts between the players are completely avoided similar to the previous work \cite{Amakasu}. Furthermore, when two photons are observed in separate output ports, the conditional probability that Player 1 chooses Arm $n$ (referred to as ``output probability" in order to distinguish it from the desired probability encoded in the OAM state) can be calculated as follows:
\begin{align}
    q_{1,n}=&\frac{1}{ p_{\mathrm{sep}}}\sum_{n_{2}=1}^{N}\mathrm{Pr}\left(n,-n_{2}\right)\nonumber\\
    \begin{split}
        =&\frac{1}{4 p_{\mathrm{sep}}}\biggl(\hat{p}_{1,n}+\hat{p}_{2,n}\\
        &  -2\sqrt{\hat{p}_{1,n}\hat{p}_{2,n}}\sum_{n^{\prime}=1}^{N}\sqrt{\hat{p}_{1,n^{\prime}}\hat{p}_{2,n^{\prime}}}\cos \left(\omega_{n^{\prime}}-\omega_{n}\right)\biggr),\label{output_prob1}
    \end{split}
\end{align}
where it is assumed that $p_{\mathrm{sep}}\neq 0$. Similarly, when calculating the output probability of Player 2, the same result as Eq. \eqref{output_prob1} is obtained. Thus, 
\begin{align}
    q_{2,n}=q_{1,n}.\label{output_equal}
\end{align}
Thus, at every step both players share the same output probabilities.
From Eq. \eqref{output_prob1}, it is clear that the output probabilities $q_{1,n}$ and $q_{2,n}$ generally differ from the desired probabilities $\hat{p}_{1,n}$ and $\hat{p}_{2,n}$.

\subsection{Optimization of the phases of OAM states}
\label{subsec:optimization}
The proposed method requires repeating the operation until two photons are observed at separate output ports, and we denoted the probability that two photons are observed at separate output ports as $p_{\mathrm{sep}}$. 
When the selection is done via relative attenuation of the signals, as in the previous work \cite{Amakasu}, $p_{\mathrm{sep}}$ is always 1/2.
On the other hand, in our new proposed method, the desired probabilities are encoded into the OAM states, which causes the value of $p_{\mathrm{sep}}$ to vary accordingly.

From Eq. \eqref{general_p_separate}, $p_{\mathrm{sep}}$ is given by
\begin{align}
    p_{\mathrm{sep}}=\frac{1}{2}-\frac{1}{2}\left|\braket{\phi_{1}}{\mathscr{R}\phi_{2}}\right|^2\label{p_sepqrate}.
\end{align}
where $\left|\braket{\phi_{1}}{\mathscr{R}\phi_{2}}\right|^2$ is the fidelity between $\ket{\phi_{1}}$ and $\ket{\mathscr{R}\phi_{2}}$.
As $p_{\mathrm{sep}}$ is monotonically decreasing with respect to $\left|\braket{\phi_{1}}{\mathscr{R}\phi_{2}}\right|^2$, it takes its maximum value of 1/2 if $\left|\braket{\phi_{1}}{\mathscr{R}\phi_{2}}\right|^2 = 0$, and it takes 0 if $\left|\braket{\phi_{1}}{\mathscr{R}\phi_{2}}\right|^2 = 1$.

$p_{\mathrm{sep}}$ must be maintained at high value for efficiency. 
It is especially crucial to avoid $p_{\mathrm{sep}}$ approaching 0, as that would mean the effective shutdown of the system. 
While the amplitudes of the OAMs are already determined through the desired probabilities $\hat{p}_{m,n}$, we can still use the phases to reduce the fidelities.
We adjust the value of the phases  $\theta_{1,1},\,\cdots,\,\theta_{1,N},\,\theta_{2,1},\,\cdots,\,\theta_{2,N}$ in order to minimize the fidelity (loss) $\mathcal{L}=\left|\braket{\phi_{1}}{\mathscr{R}\phi_{2}}\right|^2$. 
From Eqs. \eqref{OAM_1} and \eqref{OAM_2},
\begin{align}
     \begin{split}
        \mathcal{L}&=\sum_{n=1}^{N}\hat{p}_{1,n}\hat{p}_{2,n}\\
        &\quad+2\sum_{n_{1}<n_{2}}\sqrt{\hat{p}_{1,n_{1}}\hat{p}_{2,n_{1}}\hat{p}_{1,n_{2}}\hat{p}_{2,n_{2}}}\cos (\omega_{n_{2}}-\omega_{n_{1}}).
    \end{split}
    \label{fidelity}
\end{align}
Accordingly, we need to minimize $\mathcal{L}$ by changing $\omega_{1},\cdots,\omega_{N}$. 
However, because players do not share information with each other, they also do not know which phases the other player chose.
To still approximate the loss, we let each player assumes that the opponent's desired probability $\hat{p}_{m,n}$ is equal to their own desired probability.
Thus, the objective function $\mathcal{L}_{m}$ which Player $m\,(m=1,\,2)$ optimizes is as follows:
\begin{align}
    \mathcal{L}_{m} = \sum_{n=1}^{N}\hat{p}_{m,n}^{\ 2}+2\sum_{n_{1}<n_{2}}\hat{p}_{m,n_{1}}\hat{p}_{m,n_{2}}\cos (\omega_{n_{2}}-\omega_{n_{1}}).
    \label{optimization of each player}
\end{align} 
Various standard methods can be employed to solve this optimization problem by either player.
Once Player $m$ has determined $\hat{\omega}_{m,1},\cdots,\hat{\omega}_{m,N}$ as the optimal solution, the player sets values of their phases as
\begin{align}
    \theta_{m,n}=(-1)^{m}\frac{\hat{\omega}_{m,n}}{2}.
\end{align}
Consequently, the actual value of the phase differences is the averages of the optimal solution of players:
\begin{align}
    \omega_{n}=\frac{\hat{\omega}_{1,n}+\hat{\omega}_{2,n}}{2}.
\end{align}
Thus, if the estimation of players is approximately equal $(\hat{p}_{1,n}\approx\hat{p}_{2,n})$, the optimal solution of $\mathcal{L}_{1}$ and $\mathcal{L}_{2}$ are expected to be also approximately equal to the optimal solution of $\mathcal{L}$, and $\left|\braket{\phi_{1}}{\mathscr{R}\phi_{2}}\right|^2$ is expected to be minimized.

In the ``exploitation" stage, desired probabilities of players are expected to converge to identical values.
For a simple illustration, let us assume that the reward probabilities of each arm satisfy $\mu_{1}>\mu_{2}>\cdots >\mu_{N}$. 
In the ``exploitation" stage, it is expected that both players only select the best arm and the second-best arm (Arm 1 and Arm 2 in this assumption), which means $\hat{p}_{1,1},\,\hat{p}_{1,2},\,\hat{p}_{2,1},\,\hat{p}_{2,2}\approx\frac{1}{2}$, and the other values of the desired probabilities are approximately equal to zero. 
Then, from Eq. \eqref{optimization of each player}, the objective function of each player is 
\begin{align}
    \mathcal{L}_{m}\approx\frac{1}{2}\left(1+\cos \left(\omega_{2}-\omega_{1}\right)\right).
\end{align}
Obviously, we can get the minimum value $0$ when $\omega_{2}-\omega_{1}=\pi+2\pi z$, where $z$ is an arbitrary integer. 
Consequently, $p_{\mathrm{sep}}\approx 1/2$, and $q_{1,1},\,q_{1,2},\,q_{2,1},\,q_{2,2}\approx 1/2$. 
Therefore, it is expected that the output probabilities are approximately equal to the desired probabilities in the ``exploitation" stage. In this case, the optimization based on the assumption that players' estimations are equal is likely to function accurately. 

Conversely, in the ``exploration" stage, desired probabilities of players are subject to significant fluctuations. Therefore, the assumption of equivalent estimations among players is not necessarily true. However, from Eq. \eqref{fidelity}, the following inequality holds:
\begin{align}
    \left|\braket{\phi_{1}}{\mathscr{R}\phi_{2}}\right|^2&\le \sum_{n=1}^{N}\hat{p}_{1,n}\hat{p}_{2,n}+2\sum_{n_{1}<n_{2}}\sqrt{\hat{p}_{1,n_{1}}\hat{p}_{2,n_{1}}\hat{p}_{1,n_{2}}\hat{p}_{2,n_{2}}}\nonumber\\
    &= \frac{\left(\sqrt{\bm{\hat{p}}_{1}}\cdot\sqrt{\bm{\hat{p}}_{2}}\right)^2}{\left\|\sqrt{\bm{\hat{p}}_{1}}\right\|^{2}\left\|\sqrt{\bm{\hat{p}}_{2}}\right\|^2}\nonumber\\
    &= \left\{\cos \left(\sqrt{\bm{\hat{p}}_{1}},\sqrt{\bm{\hat{p}}_{2}}\right)\right\}^2,
\end{align}
where 
\begin{align}
    \sqrt{\bm{\hat{p}}_{m}} = 
    \begin{pmatrix}
        \sqrt{\hat{p}_{m,1}} \\
        \sqrt{\hat{p}_{m,2}} \\
        \vdots \\
        \sqrt{\hat{p}_{m,N}}
    \end{pmatrix}
    \quad \text{($m=1,2$)},
\end{align}
and $\cos \left(\sqrt{\bm{\hat{p}}_{1}},\sqrt{\bm{\hat{p}}_{2}}\right)$ is the cosine similarity between $\sqrt{\bm{\hat{p}}_{1}}$ and $\sqrt{\bm{\hat{p}}_{2}}$. 
This inequality means that when desired probabilities of players are different, $\left|\braket{\phi_{1}}{\mathscr{R}\phi_{2}}\right|^2$ tends to be smaller regardless of the values of phases $\hat{\omega}_{m,n}$. Consequently, it is anticipated that $\left|\braket{\phi_{1}}{\mathscr{R}\phi_{2}}\right|^2$ can be maintained at a small value during ``exploration", leading to a high value of $p_{\mathrm{sep}}$. 


\section{Numerical simulation}

\subsection{Settings of numerical simulation}\label{sec:setting}
We investigated the behavior of the proposed method through numerical simulations under two sets of environments, called Env. 1-1 and Env. 2-1.
In Env. 1-1, the number of arms is set to $N=5$, while in Env. 2-1, the number of arms is set to $N=10$. 
The reward probabilities were configured as follows:
\begin{align}
    \mu_{n}=1-\frac{n}{N+1}\ (n=1,\cdots,N).
\end{align}
Furthermore, to also examine the impact of arm index assignments, we prepared environments where the reward probabilities are rearranged. 
The specific configurations of the reward environments are detailed in TABLE \ref{table:env}. The details on configuring these reward environments are explained in Appendix \ref{appendix_setting}.
\begin{table*}[t]
  \centering
  \caption{Reward environments used in the simulation. Env. 1-1 and Env. 2-1 serve as the baseline reward environments, while the other reward environments were used to investigate the impact of arm index assignments in each method. Additionally, the bolded values indicate either the highest or the second-highest reward probability.}
  {\renewcommand\arraystretch{1.5}
  \begin{tabular}{ccc}
  \hline\hline
   & Number of arms $N$ & Reward probabilities $[\mu_{1},\mu_{2},\cdots,\mu_{N}]$\\
  \hline
   Env. 1-1& $5$ & $[\bm{\frac{5}{6}},\bm{\frac{4}{6}},\frac{3}{6},\frac{2}{6},\frac{1}{6}]$ \\
   Env. 1-2& $5$ & $[\bm{\frac{5}{6}},\frac{3}{6},\bm{\frac{4}{6}},\frac{2}{6},\frac{1}{6}]$ \\
  \hline
  Env. 2-1& $10$ & $[\bm{\frac{10}{11}},\bm{\frac{9}{11}},\frac{8}{11},\frac{7}{11},\frac{6}{11},\frac{5}{11},\frac{4}{11},\frac{3}{11},\frac{2}{11},\frac{1}{11}]$ \\
  Env. 2-2& $10$ & $[\bm{\frac{10}{11}},\bm{\frac{9}{11}},\frac{5}{11},\frac{7}{11},\frac{6}{11},\frac{8}{11},\frac{4}{11},\frac{3}{11},\frac{2}{11},\frac{1}{11}]$ \\
  Env. 2-3& $10$ & $[\bm{\frac{10}{11}},\frac{8}{11},\frac{5}{11},\frac{7}{11},\frac{6}{11},\bm{\frac{9}{11}},\frac{4}{11},\frac{3}{11},\frac{2}{11},\frac{1}{11}]$ \\
  \hline\hline
  \end{tabular}
  }
  \label{table:env}
\end{table*}

The number of selections by players, in other words, the termination time of the algorithm was set to $T=10000$, and the experiment was conducted $E=5000$ times using different random seeds (for determining the measurement outcomes and probabilistic rewards). 
When both players optimized Eq. \eqref{optimization of each player}, they used a sequential quadratic programming method as their optimization method. The initial values of $\hat{\omega}_{m,n}$ in the optimization is set to 
\begin{align}
    \hat{\omega}_{m,n}^{(0)}=\frac{2(n-1)\mathrm{\pi}}{N}.
\end{align}
The Softmax method \cite{softmax} was used to calculate the desired probabilities $\hat{p}_{m, n}$. 
$\beta_{1}(t)$ and $\beta_{2}(t)$ of the Softmax method at time step $t=1,\cdots,T$ are set according to $\beta_{1}(t)=\beta_{2}(t)=\lambda t$. However, until each arm is selected at least once, the player sets $\beta_{m}=0$ and selects the arms uniformly.

\subsection{Results}
\begin{figure}[t]
    \centering
    \includegraphics[width=0.95\linewidth,keepaspectratio]{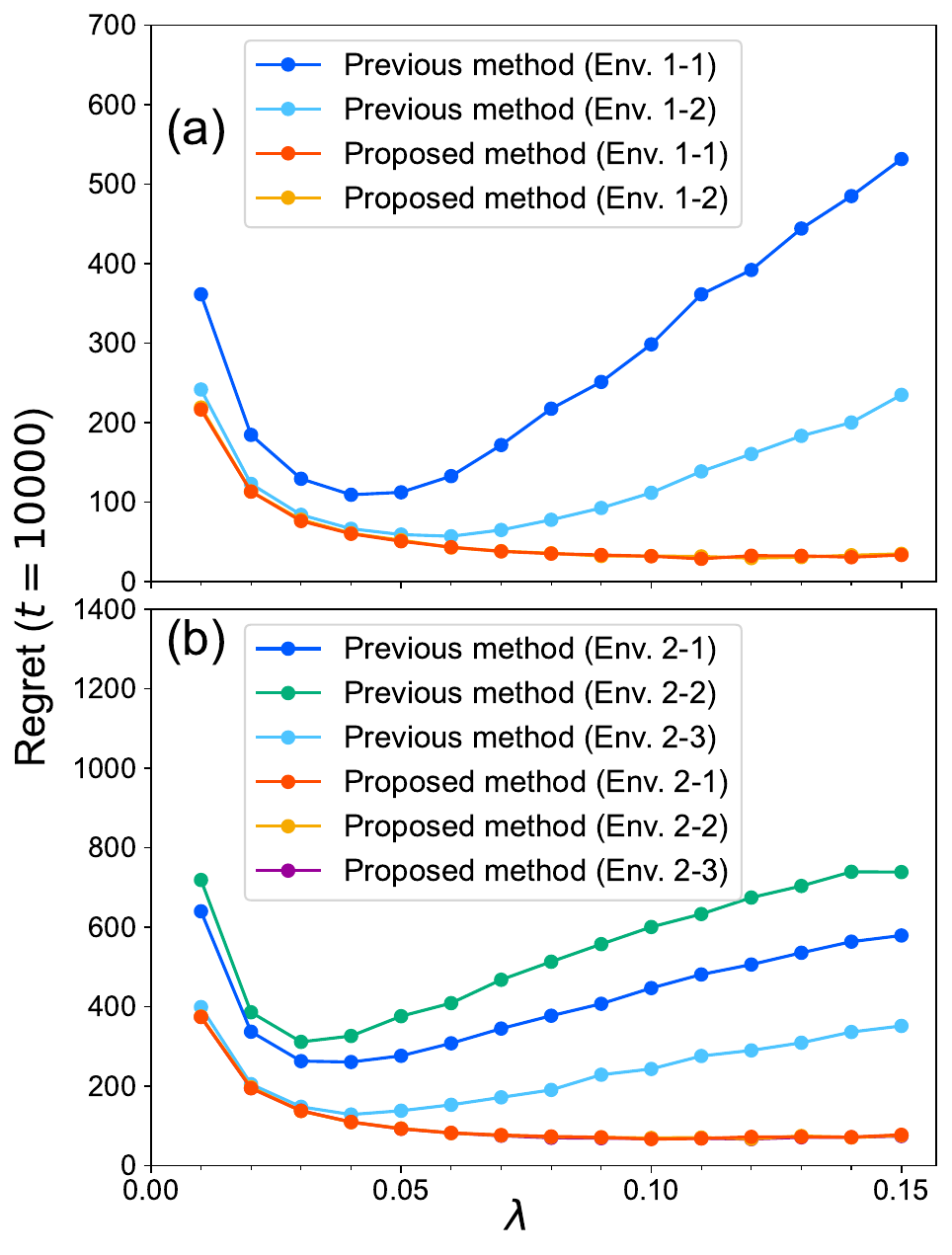}
    \caption{Comparison of the final regret values calculated from Eq.~\eqref{regret_expression} (when $t=10000$) under different settings of the hyperparameter $\lambda$ for each reward environment (see TABLE~\ref{table:env} comparing our proposed method and the previous method of Ref.~\cite{Amakasu}. (a) For reward environments Env. 1-1 and Env. 1-2, where the number of arms is $N=5$. (b) For reward environments Env. 2-1, Env. 2-2 and Env. 2-3, where the number of arms is $N=10$.}
    \label{regret_compare-fig}
\end{figure}

\begin{figure}[t]
    \centering
    \includegraphics[width=0.95\linewidth,keepaspectratio]{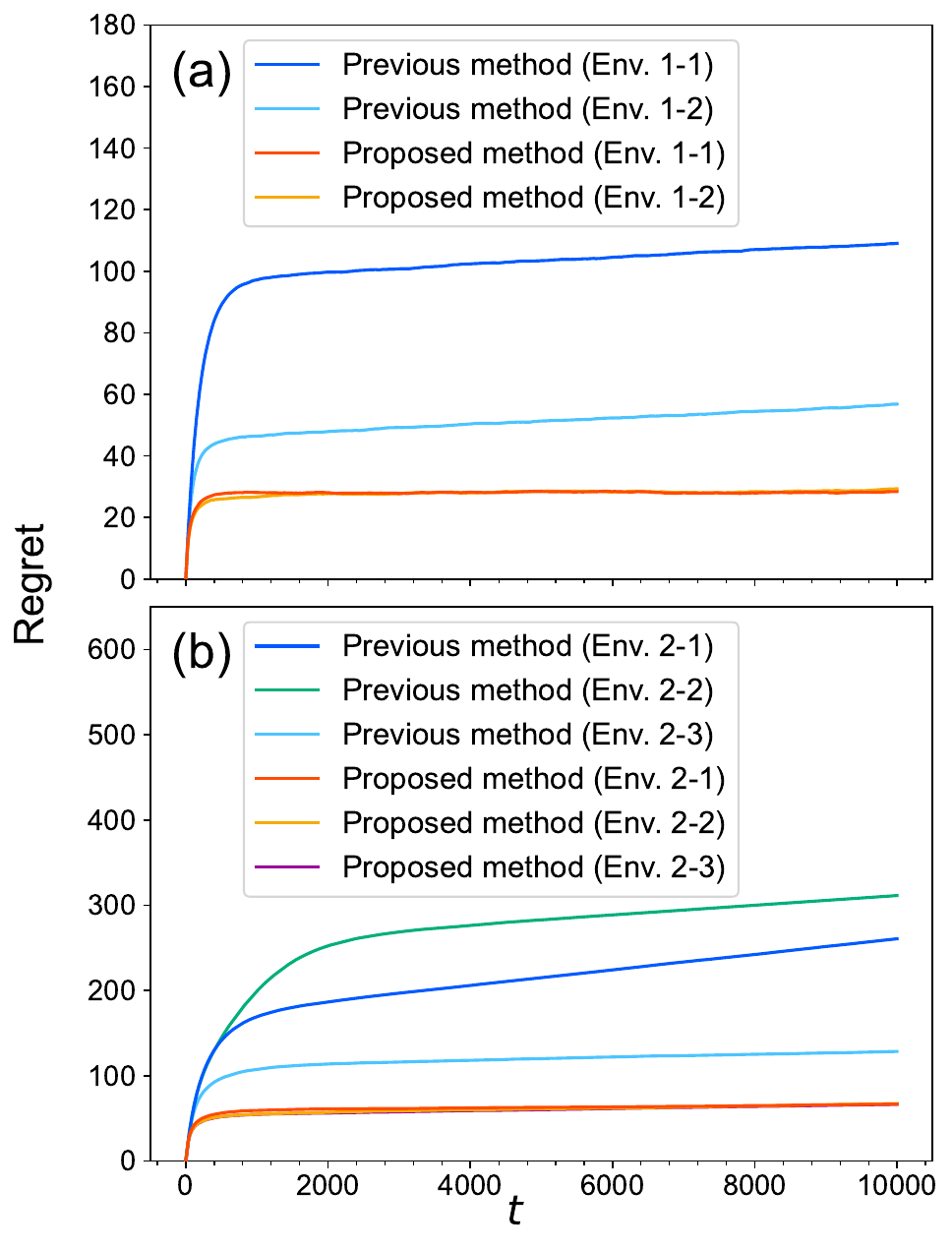}
    \caption{Comparison of regret calculated from Eq.~\eqref{regret_expression} for each reward environment (see TABLE~\ref{table:env} comparing our proposed method and the previous method of Ref.~\cite{Amakasu} when using the optimal hyperparameter $\lambda$. The optimal hyperparameter for each reward environment and method was explored in increments of 0.01 within the range of 0.01 to 0.15, as shown in FIG. \ref{regret_compare-fig}. (a) For reward environments Env. 1-1 and Env. 1-2, where the number of arms is $N=5$. (b) For reward environments Env. 2-1, Env. 2-2 and Env. 2-3, where the number of arms is $N=10$.}
    \label{regret_best_param-fig}
\end{figure}

\begin{figure}[t]
    \centering
    \includegraphics[width=\linewidth,keepaspectratio]{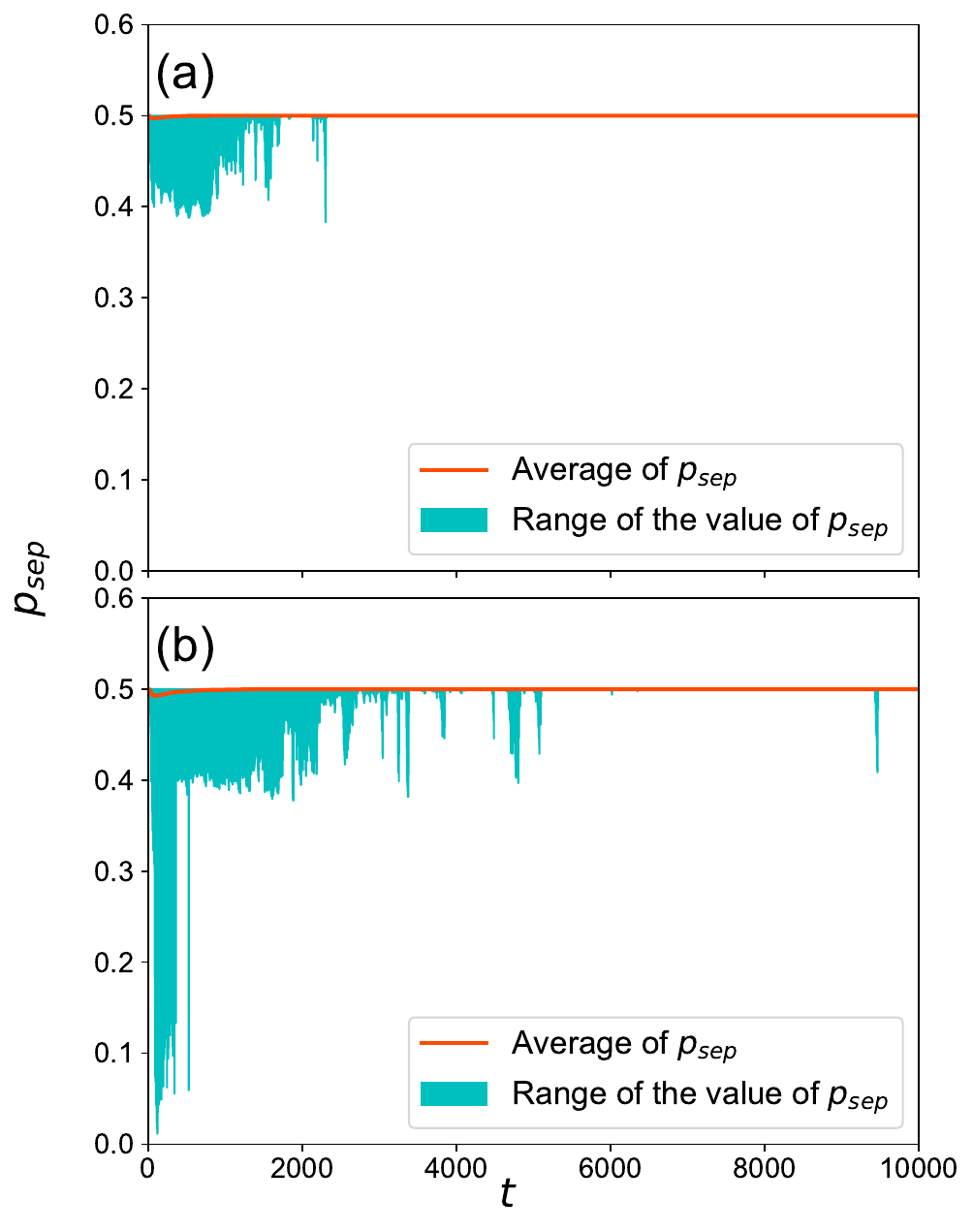}
    \caption{$p_{\mathrm{sep}}$ of the proposed method in the baseline reward environment Env. 1-1 and Env. 2-1. $\lambda$ is chosen as the optimal value that minimizes the final value of regret (In Env. 1-1, $\lambda=0.11$, and in Env. 2-1, $\lambda=0.1$). The red curve represents the average of $p_{\mathrm{sep}}$, while the cyan blue area indicates the range between the minimum and maximum values observed over 5000 executions of the algorithm. (a) Env. 1-1. (b) Env. 2-1.}
    \label{p_separate-fig}
\end{figure}

\begin{figure}[t]
    \centering
    \includegraphics[width=\linewidth,keepaspectratio]{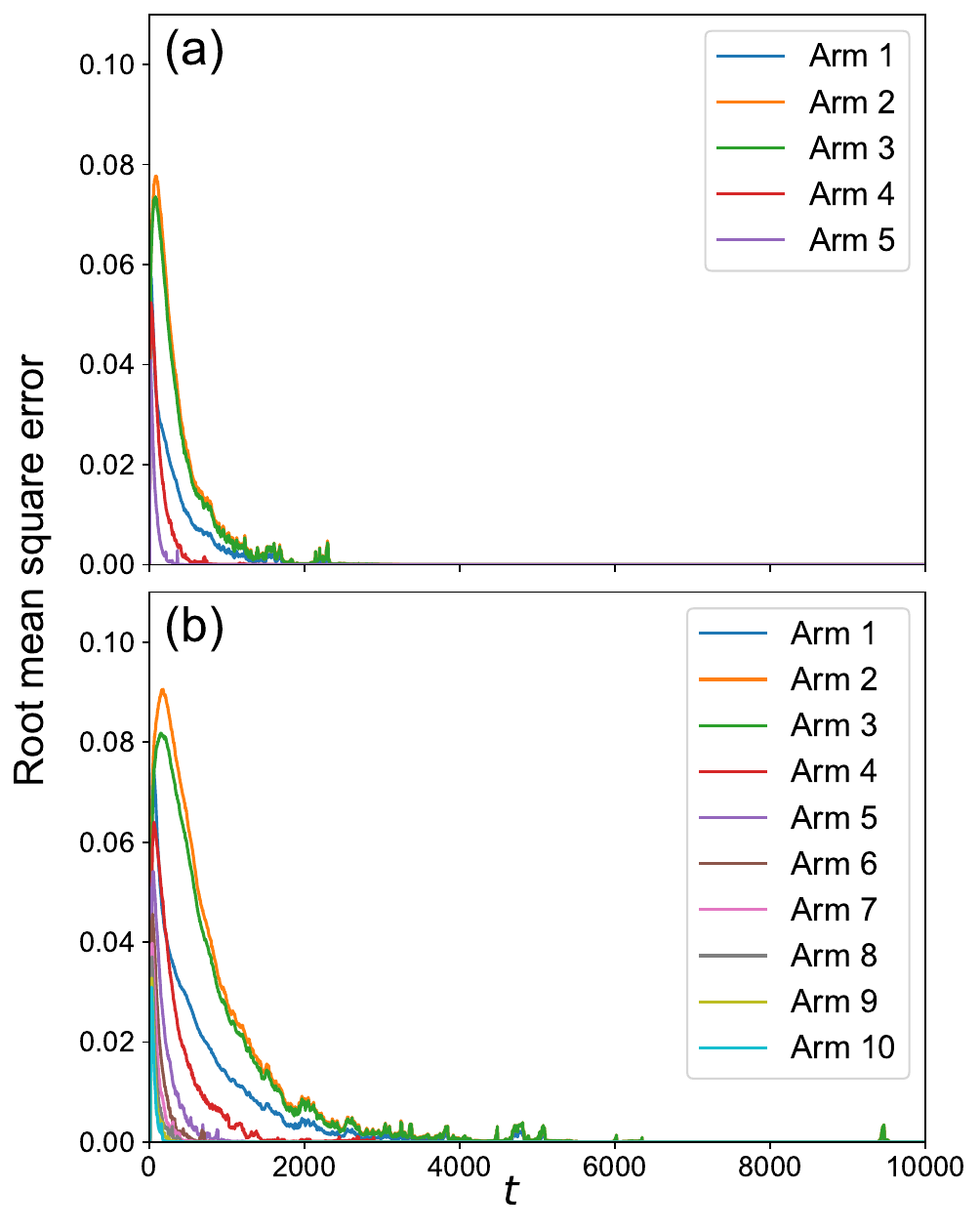}
    \caption{The Root Mean Square Error (RMSE) between desired probabilities and output probabilities for each arm of the proposed method in the baseline reward environment Env. 1-1 and Env. 2-1, averaged over players. $\lambda$ is chosen as the optimal value that minimizes the final value of regret (In Env. 1-1, $\lambda=0.11$, and in Env. 2-1, $\lambda=0.1$). (a) Env. 1-1. (b) Env. 2-1.}
    \label{rmse-fig}
\end{figure}
First, we examined the performance of the proposed method by comparing it with the previous method. The performance of each method was evaluated using the regret. The regret is defined as follows:
\begin{align}
    \begin{split}
        \mathrm{Regret}(t)&=(\mu^{*}+\mu^{**})t\\
        & \quad -\mathbb{E}\left[\sum_{\tau=1}^{t}\mathcal{C}_{n_{1}(\tau),n_{2}(\tau)}\left(\mu_{n_{1}(\tau)}+\mu_{n_{2}(\tau)}\right)\right].
    \end{split}
    \label{regret_expression}
\end{align}
Here, $\mu^{*}$ represents the highest reward probability, $\mu^{**}$ represents the second-highest reward probability, and $n_{1}(t)$ and $n_{2}(t)$ are the indices of the arms selected by each player at time $t$. Additionally, $\mathcal{C}_{i,j}$ is the indicator of the selection conflict;
\begin{align}
    \mathcal{C}_{i,j} = 1 - \frac{1}{2}\delta_{ij},
\end{align}
where $\delta_{i,j}$ is Kronecker's delta. 
The first term in Eq. \eqref{regret_expression} represents the maximum sum of cumulative reward achievable up to time $t$, while the second term denotes the expected cumulative reward sum actually obtained. 
In other words, regret quantifies the loss relative to the maximum possible cumulative reward. 
In the simulation, the second term was calculated as the empirical mean obtained from $E$ runs of each method. 

FIG. \ref{regret_compare-fig} presents the comparison of the final regret values (when $t=10000$) under different settings of the hyperparameter $\lambda$ for each reward environment and method. FIG. \ref{regret_compare-fig} (a) presents the results for Env. 1-1 and Env. 1-2, where the number of arms is $N=5$, and FIG. \ref{regret_compare-fig} (b) the results for Env. 2-1, Env. 2-2 and Env. 2-3, where the number of arms is $N=10$. 
When examining FIG. \ref{regret_compare-fig} (a) with $N=5$, it becomes evident that the previous method (based on attenuating signals) had a strong dependence on the ordering of the arms. 
In particular, it performed better in Env. 1-2, where reward probabilities are swapped compared to Env. 1-1, irrespective of the value of $\lambda$. 
In contrast, our new proposed method shows nearly identical final regret values for both Env. 1-1 and Env. 1-2 across all $\lambda$ values. 
This demonstrates that, it is not affected by the arm index allocation.

Furthermore, our new proposed method consistently outperforms the baseline method across all $\lambda$ values. 
Notably, the difference of the final regret values between the previous method and the proposed method widens as $\lambda$ increases. 
Since a larger $\lambda$ corresponds to an earlier transition from ``exploration" to ``exploitation," it can be inferred that the superior performance of our new proposed method stems from its exploration phase. 

Similar observations can be made for FIG. \ref{regret_compare-fig} (b) with $N=10$. 
The previous method performs better in environments with more favorable arm index assignments, specifically in the order Env. 2-3, Env. 2-1, and Env. 2-2. 
Conversely, our new proposed method's performance remains largely unaffected by the arm index allocation. As with $N=5$, the proposed method consistently outperforms the baseline method across all $\lambda$ values in Env. 2-3, with the performance gap becoming more pronounced at larger $\lambda$ values.

Furthermore, in both FIG. \ref{regret_compare-fig} (a) and (b), the performance of the previous method varies significantly depending on the value of $\lambda$, whereas our new proposed method exhibits relatively modest performance changes with variations in $\lambda$. 
Although not shown in FIG \ref{regret_compare-fig}, the final regret values of the proposed method remain highly stable even for $\lambda>0.15$. 
Therefore, the proposed method can be considered robust to the choice of the hyperparameter $\lambda$. 
This robustness is particularly important in the CMAB problem, where repeatedly running the algorithm to fine-tune the hyperparameter is often impractical.

FIG.~\ref{regret_best_param-fig} compares the regret for each reward environment and method using the optimal hyperparameter $\lambda$.
Here, the regret is plotted as a function of plays $t$ averaged over the $E=5000$ trials.
Our new proposed method achieves higher performance than the previous method with optimal $\lambda$. 
Analyzing the temporal changes in regret reveals that the proposed method transitions from ``exploration" to ``exploitation" more quickly than the previous method, resulting in a lower final regret value. 
The proposed method also has a smaller slope for large $t$, indicating that choices are near-perfect.

From these results, it is evident that the proposed method can solve CMAB problems with five or more arms efficiently. 
Moreover, as discussed in the previous chapter, the proposed method involves the optimization of $p_{\mathrm{sep}}$, and it is necessary to verify whether this approach is functioning appropriately.
FIG. \ref{p_separate-fig} illustrates the variation in $p_{\mathrm{sep}}$ when applying the proposed method to the baseline reward environments, Env. 1-1 ($N=5$) and Env. 2-1 ($N=10$). 
The red curve represents the mean of $p_{\mathrm{sep}}$, while the cyan blue area indicates the range between the minimum and maximum values observed across 5000 executions. 
First, in both (a) $N=5$ and (b) $N=10$, the average of $p_{\mathrm{sep}}$ remains close to its maximum value of $1/2$. This indicates that the optimization of $p_{\mathrm{sep}}$ was successful in most cases across 5000 algorithm executions. Observing the cyan blue regions, it can be seen that $p_{\mathrm{sep}}$ occasionally drops to lower values in both (a) and (b). In (a), $p_{\mathrm{sep}}$ does not drop below approximately 0.4, suggesting that the impact of such decreases is relatively minor. Conversely, in (b), there are instances where $p_{\mathrm{sep}}$ drops to significantly lower values, which can considerably reduce system efficiency. However, we confirmed that these drops are localized, and the value of $p_{\mathrm{sep}}$ quickly returns to values near its maximum of $1/2$. Additionally, in both (a) and (b), decreases in $p_{\mathrm{sep}}$ are more likely to occur when time $t$ is small, i.e., during the ``exploration" stage. This behavior can be attributed to the higher variability in desired probabilities during the ``exploration" stage, which increases the likelihood of discrepancies in desired probabilities among players. As a result, the assumption of equal desired probabilities across players, which underlies the optimization of $p_{\mathrm{sep}}$, is temporarily violated more frequently.

Based on these results, it is clear that the optimization of $p_{\mathrm{sep}}$ is functioning appropriately, and the proposed method is capable of solving the CMAB problem with two players. 
Additionally, as mentioned in the previous chapter, there is generally a discrepancy between the desired probabilities encoded into the OAM states and the actual output probabilities, which are the probabilities with which each arm is selected. Therefore, as further evidence, the difference between the desired probabilities and the output probabilities for each arm was evaluated using the Root Mean Square Error (RMSE). 
The RMSE is defined as follows:
\begin{align}
    \mathrm{RMSE}(t,n)=\frac{1}{2}\sum_{m=1}^{2}\sqrt{\frac{1}{E}\sum_{e=1}^{E}(\hat{p}_{m,n}^{(e)}(t)-q_{m,n}^{(e)}(t))^{2}},
\end{align}
where $\hat{p}_{m,n}^{(e)}(t)$ represents the desired probability during the $e$-th execution ($1\leq e\leq E$), and $q_{m,n}^{(e)}(t)$ denotes the output probability during the $e$-th execution.
FIG. \ref{rmse-fig} illustrates the RMSE between desired probabilities and output probabilities. In (a) $N=5$ and (b) $N=10$, differences between desired probabilities and output probabilities are observed during the ``exploration" stage. However, these differences approach zero as the algorithm transitions toward the ``exploitation" stage.
The average RMSE of these differences remains at most approximately 0.08, suggesting that the impact on the algorithm's performance is not critical. Nevertheless, it is possible that these discrepancies between desired probabilities and output probabilities might be avoidable and reducing the efficiency of ``exploration," warranting further investigation.

\section{Conclusion}
This paper proposes a novel conflict-free approach to solve the competitive multi-armed bandit problem with two players and an arbitrary number of arms. 
This scalability was achieved by exploiting the quantum properties of photons. 
In previous works, attenuators were used to reflect desired probabilities. 
However, the new proposed method presented here encodes desired probabilities into the amplitudes of the photon states. 
We derived how these amplitudes can be modified after observations in a reinforcement-learning approach based on the softmax method.
We then examined the rate of successful separation of the photons and the stability of the procedure over many plays and with varying meta-parameters.
Numerical simulations demonstrated that the proposed method outperformed the previous approach under all conditions.

Future research needs to focus on further improving the method proposed in this paper. 
Potential approaches include optimizing $p_{\mathrm{sep}}$ more rapidly and developing techniques to prevent sudden fluctuations in $p_{\mathrm{sep}}$ that occur infrequently. 
This might be achievable by slightly relaxing the no-communication condition between players. 
Additionally, for understanding the impact of discrepancies between desired probabilities and output probabilities on performance or universal performance evaluations across various reward environments, more detailed theoretical analyses are needed. 
Furthermore, it is essential to consider collective conflict-free decision making systems capable of addressing the CMAB problem involving three or more players.

Our successful demonstration shows how fundamental aspects of quantum optics can be exploited for solving problems from reinforcement learning. 
Although the field is not yet sufficiently developed for practical real-world application, our method relies on simple rules that may be realizable directly in specialized hardware, without the need of classical computers.
Building such a repertoire of functionality for a variety of model tasks is a promising route towards the development of specialized hardware based on quantum optical principles, and is likely to remain a promising research area for the foreseeable future.

\begin{acknowledgments}
This research was funded in part by the Japan Society for the Promotion of Science through Grant-in-Aid for Transformative Research Areas (A) (JP22H05197) and the Japan Science and Technology Agency SPRING (JPMJSP2108).
\end{acknowledgments}

\appendix

\section{Transformation of orbital angular momentum states}\label{Appendix_beamsplitter}

In this section, we derive Eqs. \eqref{OAM_trans_1} and \eqref{OAM_trans_2}, which describe the transformation of the input OAM state of photons. When a photon is reflected by a mirror or a beamsplitter, its OAM is inverted. Consequently, the transformation of the OAM state induced by a mirror or beamsplitter is generally expressed as follows:
\begin{align}
    \ket{\phi} &\xrightarrow{\mathrm{Mirror}} \mathrm{i}\mathscr{R}\ket{\phi}\\
    \ket{\phi} &\xrightarrow{\mathrm{Beam splitter}} \frac{1}{\sqrt{2}}\ket{\phi}_{\mathrm{transmitted}}+\frac{\mathrm{i}}{\sqrt{2}}\mathscr{R}\ket{\phi}_{\mathrm{reflected}}
\end{align}
where $\mathscr{R}$ represents an operator that inverts the OAM, expressed as Eq. \eqref{reflection}, and $\ket{\phi}_{\mathrm{transmitted}}$ and $\ket{\phi}_{\mathrm{reflected}}$ denote the OAM states of the photon upon transmission through and reflection by the beamsplitter, respectively.
As shown in FIG. \ref{proposed-fig}, in the proposed system, the input OAM states of the two photons are transformed as follows:
\begin{align}
    \ket{\phi_{1},1}&\xrightarrow{\mathrm{Beam splitter}}\frac{\mathrm{i}}{\sqrt{2}}\ket{\mathscr{R}\phi_{1},1}+\frac{1}{\sqrt{2}}\ket{\phi_{1},2}\nonumber\\
    &\xrightarrow{\mathrm{Mirror}}\frac{\mathrm{i}^2}{\sqrt{2}}\ket{\mathscr{R}^{2}\phi_{1},1}+\frac{\mathrm{i}}{\sqrt{2}}\ket{\mathscr{R}\phi_{1},2}\nonumber\\
    &= -\frac{1}{\sqrt{2}}\ket{\phi_{1},1}+\frac{\mathrm{i}}{\sqrt{2}}\ket{\mathscr{R}\phi_{1},2}\label{derive_trans_1}\\
    \ket{\phi_{2},2}&\xrightarrow{\mathrm{Beam splitter}}\frac{1}{\sqrt{2}}\ket{\phi_{2},1}+\frac{\mathrm{i}}{\sqrt{2}}\ket{\mathscr{R}\phi_{2},2}\nonumber\\
    &\xrightarrow{\mathrm{Mirror}}\frac{\mathrm{i}}{\sqrt{2}}\ket{\mathscr{R}\phi_{2},1}+\frac{\mathrm{i}^2}{\sqrt{2}}\ket{\mathscr{R}^{2}\phi_{2},2}\nonumber\\
    &=\frac{\mathrm{i}}{\sqrt{2}}\ket{\mathscr{R}\phi_{2},1}-\frac{1}{\sqrt{2}}\ket{\phi_{2},2}\label{derive_trans_2}.
\end{align}
Eqs. \eqref{derive_trans_1} and \eqref{derive_trans_2} correspond to Eqs. \eqref{OAM_trans_1} and \eqref{OAM_trans_2}.

\section{Details on the configuration of the reward environments in the numerical simulation}\label{appendix_setting}

In Sec. \ref{sec:setting}, the five types of reward environments used in the numerical simulations are described. Two baseline environments, Env. 1-1 and Env. 2-1, are defined, while the other environments are created by rearranging the reward probabilities of these baseline environments. This section explains the method for configuring the rearranged environments based on the baseline environments. First, in the previous method, the probability $\mathrm{Pr}(n_{1},-n_{2})$ that Player 1 selects Arm $n_{1}$ and Player 2 selects Arm $n_{2}$ is expressed as follows \cite{Amakasu}:
\begin{align}
    \mathrm{Pr}(n_{1},-n_{2})\propto\frac{1}{N^2}\sin ^2 \left(\frac{(n_{2}-n_{1})\mathrm{\pi}}{N}\right).\label{previous_bothplayers_outputprobs}
\end{align}
Therefore, the closer $|n_{2}-n_{1}|$ is to $\frac{N}{2}$, the more likely the pair $(n_{1},n_{2})$ is to be selected, when $n_{1}\neq n_{2}$. When the index of the best, the second-best, the third-best arms are denoted as $n^{*}$, $n^{**}$, $n^{***}$, respectively, the closer $|n^{**}-n^{*}|$ is to $\frac{N}{2}$ and the farther $|n^{***}-n^{*}|$ is to $\frac{N}{2}$, the more advantageous it becomes for the previous method since distinguishing between the second-best and third-best arms is crucial in the CMAB problem with two players. Thus, Env. 1-1 can be considered a disadvantageous setting for the previous method. To create a more advantageous setting, we also considered Env. 1-2, which was obtained by swapping the values of $\mu_{2}$ and $\mu_{3}$. For Env. 2-1, we also considered two variations: Env. 2-2, which is disadvantageous for the previous method, and Env. 2-3, which is advantageous. Env. 2-2 is derived by swapping the values of $\mu_{3}$ and $\mu_{6}$ in Env. 2-1. Env. 2-3 is created by first swapping the values of $\mu_{2}$ and $\mu_{3}$ in Env. 2-1, followed by swapping the values of $\mu_{3}$ and $\mu_{6}$.

\bibliographystyle{ieeetr}
\bibliography{Konaka_oam}

\begin{thebibliography}{10}

\bibitem{sutton1999reinforcement}
R.~S. Sutton and A.~G. Barto, ``Reinforcement learning: An introduction,'' {\em Robotica}, vol.~17, no.~2, pp.~229--235, 1999.

\bibitem{MAB}
H.~Robbins, ``Some aspects of the sequential design of experiments,'' {\em Bulletin of the American Mathematical Society}, vol.~58, pp.~527--535, 1952.

\bibitem{Exploration_and_exploitation}
J.~G. March, ``Exploration and exploitation in organizational learning,'' {\em Organization Science}, vol.~2, no.~1, pp.~71--87, 1991.

\bibitem{advertisement1}
T.~Lu, D.~P{\'a}l, and M.~P{\'a}l, ``Contextual multi-armed bandits,'' in {\em Proceedings of the Thirteenth international conference on Artificial Intelligence and Statistics}, pp.~485--492, 2010.

\bibitem{advertisement2}
D.~Chakrabarti, R.~Kumar, F.~Radlinski, and E.~Upfal, ``Mortal multi-armed bandits,'' in {\em Neural Information Processing Systems}, vol.~21, 2008.

\bibitem{drug1}
W.~R. Thompson, ``On the likelihood that one unknown probability exceeds another in view of the evidence of two samples,'' {\em Biometrika}, vol.~25, no.~3-4, pp.~285--294, 1933.

\bibitem{drug2}
S.~S. Villar, J.~Bowden, and J.~Wason, ``Multi-armed bandit models for the optimal design of clinical trials: benefits and challenges,'' {\em Statistical science: a review journal of the Institute of Mathematical Statistics}, vol.~30, no.~2, p.~199, 2015.

\bibitem{drug3}
J.~C. Gittins and D.~M. Jones, ``A dynamic allocation index for the discounted multiarmed bandit problem,'' {\em Biometrika}, vol.~66, no.~3, pp.~561--565, 1979.

\bibitem{softmax}
J.~Vermorel and M.~Mohri, ``Multi-armed bandit algorithms and empirical evaluation,'' in {\em European conference on machine learning}, pp.~437--448, Springer, 2005.

\bibitem{Thompson_sampling}
O.~Chapelle and L.~Li, ``An empirical evaluation of thompson sampling,'' {\em Advances in neural information processing systems}, vol.~24, 2011.

\bibitem{UCB}
P.~Auer, N.~Cesa-Bianchi, and P.~Fischer, ``Finite-time analysis of the multiarmed bandit problem,'' {\em Machine learning}, vol.~47, pp.~235--256, 2002.

\bibitem{CMAB1}
L.~Lai, H.~El~Gamal, H.~Jiang, and H.~V. Poor, ``Cognitive medium access: Exploration, exploitation, and competition,'' {\em IEEE Transactions on Mobile Computing}, vol.~10, no.~2, pp.~239--253, 2011.

\bibitem{CMAB2}
S.-J. Kim, M.~Naruse, and M.~Aono, ``Harnessing the computational power of fluids for optimization of collective decision making,'' {\em Philosophies}, vol.~1, no.~3, pp.~245--260, 2016.

\bibitem{CMAB3}
L.~Besson and E.~Kaufmann, ``Multi-player bandits revisited,'' in {\em Algorithmic Learning Theory}, pp.~56--92, PMLR, 2018.

\bibitem{CMAB4}
S.~Maghsudi and E.~Hossain, ``Multi-armed bandits with application to 5{G} small cells,'' {\em IEEE Wireless Communications}, vol.~23, no.~3, pp.~64--73, 2016.

\bibitem{cognitive_radio}
J.~Mitola and G.~Q. Maguire, ``Cognitive radio: making software radios more personal,'' {\em IEEE personal communications}, vol.~6, no.~4, pp.~13--18, 1999.

\bibitem{OAM}
L.~Allen, M.~W. Beijersbergen, R.~J.~C. Spreeuw, and J.~P. Woerdman, ``Orbital angular momentum of light and the transformation of laguerre-gaussian laser modes,'' {\em Physical review A}, vol.~45, pp.~8185--8189, 1992.

\bibitem{Amakasu}
T.~Amakasu, N.~Chauvet, G.~Bachelier, S.~Huant, R.~Horisaki, and M.~Naruse, ``Conflict-free collective stochastic decision making by orbital angular momentum of photons through quantum interference,'' {\em Scientific Reports}, vol.~11, no.~1, p.~21117, 2021.

\bibitem{lavery2013efficient}
M.~P. Lavery, D.~J. Robertson, A.~Sponselli, J.~Courtial, N.~K. Steinhoff, G.~A. Tyler, A.~E. Willner, and M.~J. Padgett, ``Efficient measurement of an optical orbital-angular-momentum spectrum comprising more than 50 states,'' {\em New Journal of Physics}, vol.~15, no.~1, p.~013024, 2013.

\bibitem{laser1}
H.~Ito, T.~Mihana, R.~Horisaki, and M.~Naruse, ``Conflict-free joint decision by lag and zero-lag synchronization in laser network,'' {\em Scientific Reports}, vol.~14, no.~1, p.~4355, 2024.

\bibitem{laser2}
S.~Kotoku, T.~Mihana, A.~R{\"o}hm, R.~Horisaki, and M.~Naruse, ``Asymmetric leader-laggard cluster synchronization for collective decision-making with laser network,'' {\em Optics Express}, vol.~32, no.~8, pp.~14300--14320, 2024.

\bibitem{softmax_parameter}
N.~Cesa-Bianchi and P.~Fischer, ``Finite-time regret bounds for the multiarmed bandit problem,'' in {\em International Conference on Machine Learning}, 1998.

\end{thebibliography}

\end{document}